\documentclass[epj]{svjour}

\usepackage{graphics}

\begin{document}

\title{Synchronization and structure in an adaptive oscillator network}

\author{Pablo M. Gleiser\inst{1} \and Dami\'an H. Zanette\inst{2}}

\institute{Consejo Nacional de Investigaciones Cient\'{\i}ficas y
T\'ecnicas \\ Centro At\'omico Bariloche and Instituto Balseiro,
8400 Bariloche, Argentina}

\date{Received: date / Revised version: date}

\abstract{We analyze the interplay of synchronization and structure
evolution in an evolving network of phase oscillators. An initially
random network is adaptively rewired according to the dynamical
coherence of the oscillators, in order to enhance their mutual
synchronization. We show that the evolving network reaches a
small-world structure. Its clustering coefficient attains a maximum
for an intermediate intensity of the coupling between oscillators,
where a rich  diversity of synchronized oscillator groups is
observed. In the stationary state, these synchronized groups are
directly associated with network clusters. }

\maketitle

\section{Introduction}
\label{intro}

The emergence of coherent dynamics in ensembles of interacting
active elements is one of the basic manifestations of
self-organization in complex systems. A large number of natural
phenomena can be ascribed to the organized action of individual
agents, whose joint function gives rise to collective signals.
Specifically, periodic dynamics in living organisms at many levels
(molecular, cellular, social) have early been suggested to originate
in the synchronization of many elementary oscillations
\cite{Wiener,Winfree}.

The most significant instances of synchronization in natural systems
are found in the realm of life sciences. Synchronized dynamics is
essential to the function of cellular tissues such as in the brain
\cite{Gray} and in the heart \cite{Strogatz}, and is known to
characterize certain forms of behaviour in social animals, from
insects to humans \cite{Buck,Buck2,Strogatz2}. The observation that
synchronization is ubiquitous in the functioning of biological
systems immediately rises the question on the evolutionary
mechanisms that may have given origin to this form of
self-organization. Several models have been explored where dynamical
parameters are modified in response to ``selection pressure'' in the
form of learning algorithms, in such a way that the function of the
system evolves towards a specified goal \cite{Moyano,Ipsen}.

In particular, Gong and Van Leeuwen have considered an ensemble of
attractively coupled chaotic maps whose interaction pattern,
described by means of a network, evolves in such a way that the
mutual synchronization of individual motions is enhanced
\cite{Gong}. The algorithm favours the interaction between elements
whose internal states are similar. In the present paper, we consider
a similar form of adaptive evolution in a network of coupled
non-identical phase oscillators \cite{Kuramoto}. In this kind of
system, the heterogeneity of the ensemble competes with coupling
against the emergence of coherent dynamics \cite{Manrubia}. We show
that the evolution of the interaction network makes it possible to
partially synchronize groups of oscillators, for coupling
intensities well below the synchronization threshold of globally
coupled ensembles. Correspondingly, the network evolves towards a
structure with relatively high clustering, approaching a pattern
with small-world properties --as also found to occur for networks of
chaotic maps \cite{Gong}. In contrast with the latter system,
however, in oscillator ensembles clustering turns out to be maximal
for an intermediate value of the interaction strength. We
characterize the emerging dynamical and structural properties of the
ensemble as functions of the coupling intensity.

\section{The model}
\label{dynamics}

Our model consists of an ensemble of $N$ coupled phase
oscillators, whose individual evolution is given by
\begin{equation}
\dot{\phi_i} = \omega_i + \frac{r}{M_i} \sum_{j =1}^N W_{ij}
\sin{(\phi_j - \phi_i)}, \label{eq1}
\end{equation}
$i=1,\dots , N$, where $\omega_i$ is the natural frequency of
oscillator $i$ and $r$ is the coupling strength. The weights
$W_{ij}$ define the adjacency matrix of the interaction network:
$W_{ij}=1$ if oscillator $i$ interacts with oscillator $j$, and $0$
otherwise. The number of neighbours of oscillator $i$ is $M_i=\sum_j
W_{ij}$. Interactions are symmetric, so that $W_{ij}=W_{ji}$ and the
network is a non-directed graph. It is assumed that this network is
not disconnected. As explained in the following, in our model the
interaction network changes with time.

During the evolution, the network is quenched over time intervals of
length $T$. Along each one of these intervals we calculate the
average oscillation frequency of each oscillator,
\begin{equation}
\Omega_i = \frac{1}{T} \int_t^{t+T} \dot{\phi_i}(t') dt'.
\label{eq2}
\end{equation}
It is well known that, if the coupling constant $r$ is
sufficiently large, two oscillators $i$ and $j$ whose natural
frequencies are close enough will asymptotically oscillate with
the same average frequency, $\Omega_i=\Omega_j$ for $T\to \infty$.
The collective manifestation of this long-term correlation between
the dynamics of oscillator pairs belongs, precisely, to the class
of synchronization phenomena addressed to in the Introduction.

In our model, after each interval of length $T$ has elapsed, the
following adaptive mechanism is applied to make the network
structure evolve. An oscillator $i$ is chosen at random, and the
values $\delta_{ij} = |\Omega_i - \Omega_j|$ are calculated for all
$j\neq i$. We detect the oscillator $j_1$ for which  $\delta_{ij_1}$
is minimum amongst all the $\delta_{ij}$. We also detect, amongst
the neighbours of $i$, the oscillator $j_2$ for which
$\delta_{ij_2}$ is maximal. If $j_1$ is one of the neighbours of
oscillator $i$, the network is not changed. Otherwise, the network
link between $i$ and $j_2$ is replaced by a link between $i$ an
$j_1$. After this update, a new interval of length $T$ begins, and
the process is successively repeated until some kind of stationary
state is reached.

To avoid the unnatural situation where the average frequencies are
compared with too high (machine) precision, we apply the above
update mechanism only when the involved quantities $\delta_{ij_1}$
and $\delta_{ij_2}$ are larger than a certain threshold $\epsilon$.
In particular, if the maximal difference between the average
frequencies of oscillator $i$ and all its neighbours is below
$\epsilon$, no changes are made.

The mechanism by which the interaction network is rewired has been
designed as to favour the connection between oscillators with
similar average frequencies. In other words, for a given coupling
intensity, its main effect on the collective dynamics of the
oscillator ensemble is to enhance the possibility of
synchronization. This aspect is studied in Section \ref{synchr}.
Now, what is the effect on the structure of the interaction network?
Which topological features in the connection pattern emerge from
such mechanism? To answer these questions we analyze statistical
properties of the network connectivity, as described in Section
\ref{clus}.

\section{Numerical results}

In the numerical calculations, we consider an oscillator ensemble of
size $N=100$. The natural frequencies $\omega_i$ are chosen at
random from a Gaussian distribution with zero mean and unitary
variance, $g(\omega)= \exp{(-\omega^2/2)}/ \sqrt{2\pi}$. Average
frequencies are discerned with a threshold $\epsilon=10^{-3}$.
Initially, the neighbours of each oscillator are chosen at random
from the whole ensemble, establishing a link between each oscillator
pair with probability $p=0.12$. For $N=100$, this implies that each
oscillator has some $12$ neighbours on the average, which insures
that essentially in all realizations of the initial condition the
interaction network is not disconnected. As the same time, the
connections are rather sparse. The initial phases $\phi_i(0)$ are
drawn from a uniform distribution in $[0,2\pi)$. The evolution
equations are integrated using a standard Euler scheme, with time
step $\Delta t = 10^{-2}$. A rewiring of the network is attempted
every $10^3$ time steps, so that $T=10$.

\subsection{Synchronization properties} \label{synchr}

\begin{figure}
\resizebox{\columnwidth}{!}{\includegraphics*{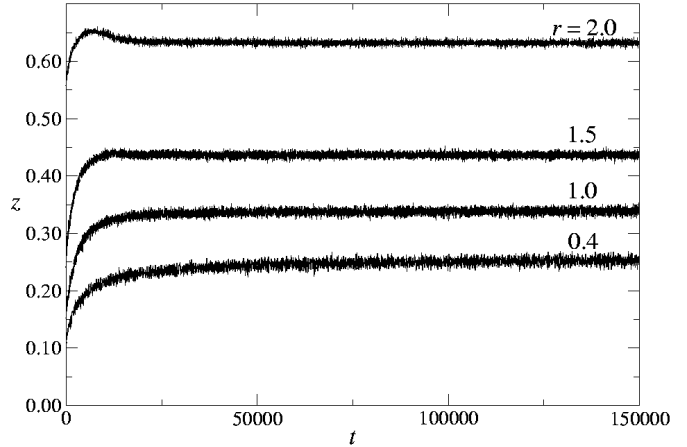}}
\caption{ Order parameter  $z$  versus time $t$,  for four different
 values of the coupling strength $r$. From bottom to top  $r=0.4$, $1.0$, $1.5$ and $2.0$.}
\label{figure1}
\end{figure}

As a first step to characterize the interplay between the collective
dynamics of the oscillators and the evolution of their interaction
network, we study the synchronization properties of the ensemble. We
begin our analysis considering the synchronization order parameter
\cite{Kuramoto}
\begin{equation}
Z(t) = \frac{1}{N}\left| \sum_{j=1}^N e^{i \phi_j(t)} \right| .
\end{equation}
The time average of $Z(t)$,
\begin{equation}
z = \frac{1}{T} \int_t^{t+T} Z (t') dt',
\end{equation}
is  calculated over intervals of length $T$, when the network is
quenched, in the same way as the average oscillation frequencies,
Eq. (\ref{eq2}). The order parameter ranges from $z \sim N^{-1/2}$
for unsynchronized motion, to $z \sim 1$ when the oscillators become
fully synchronized. In Fig. \ref{figure1} we show the time evolution
of $z$ for four different values of the coupling strength, $r=0.4$,
$1.0$, $1.5$ and $2.0$, averaged over $100$ initial conditions. In
all cases, the order parameter reaches a stationary value which
grows as the coupling strength increases.

\begin{figure}
\resizebox{\columnwidth}{!}{\includegraphics*{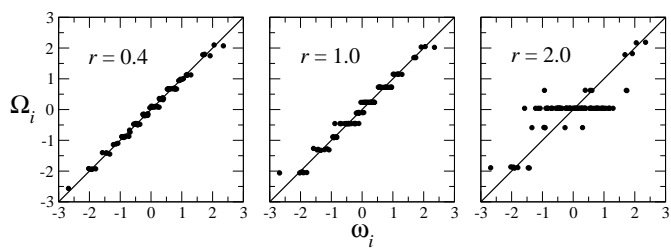}}
\caption{ Average frequency $\Omega_i$  versus natural frequency
$\omega_i$, for three different values of the coupling strength $r$.
Each dot represents an oscillator in an ensemble of size $N=100$.
The three plots are snapshots at time $t=5\times 10^4$. Straight
lines are the diagonals $\Omega_i=\omega_i$. } \label{figure2}
\end{figure}

The analysis of $z$ reveals the existence of partially synchronized
states, even for  very small values of the coupling strength. In
order to obtain further information on these states we compare the
distribution of the average frequencies $\Omega_i$ with that of the
natural frequencies $\omega_i$ for  representative single
realizations. Figure \ref{figure2} shows snapshots of the average
frequency of each oscillator as a function of its natural frequency
for three values of the coupling strength, $r=0.4$, $1.0$, and
$2.0$. For comparison, we recall that in a globally coupled ensemble
--where $W_{ij}=1$ for all $i$ and $j$, and $M_i=N$-- the
synchronization threshold with the same distribution of natural
frequencies is placed at $r_c \approx 1.6$. The three snapshots are
taken at time $t=5\times 10^4 = 50 N T$, when $z$ has practically
reached its stationary level. At this time, on the average, each
oscillator has been chosen $50$ times at the updates of the network
structure.

In the plots of $\Omega_i$ versus $\omega_i$, synchronization is
revealed by the presence of horizontal arrays of dots, corresponding
to oscillators with different natural frequencies which, through the
effect of coupling, have attained the same average frequency.
Already for $r=0.4$, we note the formation of many small groups of
synchronized oscillators, all over the frequency distribution. A
number of oscillators, however, do not belong to any group. For
$r=1.0$ several groups have mutually collapsed, and the resulting
aggregates are larger. Finally, just a few groups remain for
$r=2.0$, containing essentially all the oscillators.

In our system, thus, synchronized groups are already observed for
coupling intensities well below the synchronization threshold $r_c$
of a globally coupled ensemble, quoted above. This is an indication
that the evolution of the network structure succeeds at creating and
maintaining connections between oscillators which are more likely to
become synchronized, even for low coupling intensities. In a
globally coupled ensemble --and, more generally, in a randomly
connected ensemble with moderate to high connectivity--
synchronization would first involve those oscillators in the centre
of the Gaussian distribution of natural frequencies, where their
number is larger \cite{Kuramoto,Manrubia}. This would typically give
rise to a single synchronized aggregate around the mean natural
frequency, surrounded by a ``cloud'' of non-synchronized
oscillators. Increasing the coupling strength, the aggregate would
grow in size at the expense of the ``cloud''. In our system, on the
other hand, the appearance of small synchronized groups for low
interaction strength reveals a non-trivial structure in the
underlying network. Due to the presence of several groups, we may
expect that the collective behaviour of the ensemble is dynamically
richer than in the cases where a single aggregate forms.

\subsection{Network structure} \label{clus}

The synchronization properties discussed above --in particular, the
formation of synchronized groups for low coupling intensities--
suggests that a reasonable choice to quantitatively characterize the
structure of the interaction network is the clustering coefficient
$C$ \cite{Watts}. We recall that the clustering coefficient is a
topological property of a network which measures the average number
of neighbours of a given node which are in turn mutual neighbours.
It is defined as the ratio of the total number of triangles to the
total number of connected triples in the network. In a recent work
\cite{McGraw}, it was found that both in random and in scale-free
networks, an increase in the clustering coefficient favors the
formation of oscillator subpopulations  synchronized at different
frequencies.

\begin{figure}
\resizebox{\columnwidth}{!}{\includegraphics*{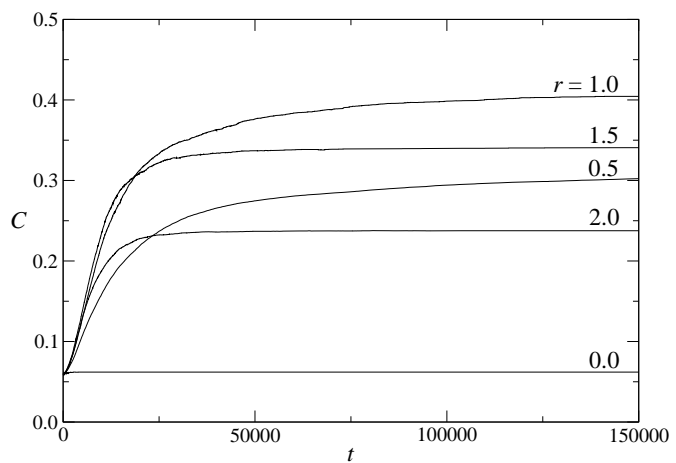}}
\caption{ Time evolution of the clustering $C$ for five different
values of the coupling strength $r$. } \label{figure3}
\end{figure}

In our model, as the network evolves, $C$ changes with time. Figure
\ref{figure3} shows the time evolution of the clustering coefficient
for several values of the coupling strength. Each curve is an
average over $20$ realizations. Even for $r=0$, in the absence of
interaction, $C$ shows a small (but fast) growth from its initial
level ($C_0 \approx 0.058$). In this case, the rewiring mechanism
tends to cluster those oscillators whose natural frequencies are
close to each other. For $r>0$, however, the effect is much
stronger. The clustering coefficient attains a long-time asymptotic
value $C_\infty$ which depends on the coupling strength $r$, and can
reach up to seven times the initial value.

\begin{figure}
\resizebox{\columnwidth}{!}{\includegraphics*{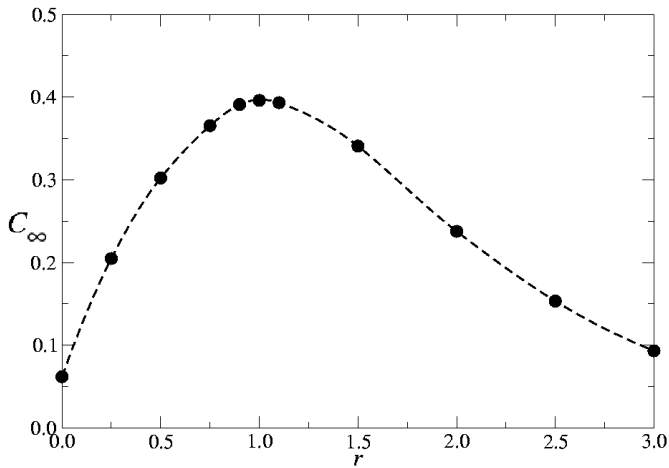}}
\caption{Stationary value of the clustering $C_{\infty}$ as a
function of the coupling strength $r$ (dots). The dashed curve is a
spline approximation added for clarity.} \label{figure4}
\end{figure}

The asymptotic clustering coefficient as a function of the coupling
strength is shown in Fig. \ref{figure4}. Interestingly enough, it
exhibits a maximum at $r_{\max} \approx 1$, where it reaches the
value $C\approx 0.4$. For $r>r_{\max}$, $C_\infty$ declines, and
seems to approach the value observed for $r=0$ for sufficiently
large coupling intensities. The presence of a maximum in $C_\infty$
for an intermediate value of $r$ indicates that, as the result of
the rewiring process, the resulting network acquires a more complex
(less random) structure when interactions between oscillators are
neither too weak nor too strong. When $r$ is small, after a few
rewiring events which connect oscillators with similar natural
frequencies --thus slightly increasing the clustering coefficient--
the network evolution ceases. Due to the modest effect of
synchronization at such low interaction intensities, there is no
much advantage, with respect to the adaptive rewiring mechanism, in
further increasing $C$. On the other hand, for large coupling
intensities, a high synchronization level is rapidly attained even
for random interaction patterns. Again, consequently, rewiring has
little effect on the network structure. The complexity of the
resulting structure is therefore maximal for moderate interaction
strengths, where the emergence of coherent synchronized oscillations
is already significant, but such that the collective dynamics is not
too much organized as to make the network evolution unnecessary.

\begin{figure}
\resizebox{\columnwidth}{!}{\includegraphics*{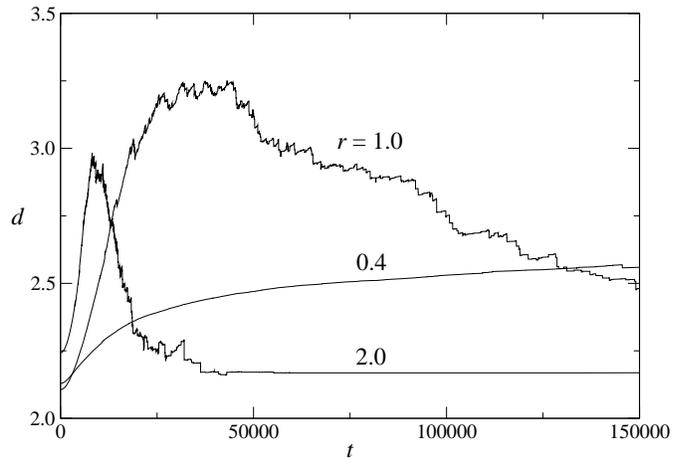}}
\caption{Mean distance $d$ as a function of time, for three
different values of the coupling strength $r$.} \label{figure5}
\end{figure}

As a next step in the analysis of the network structure, we consider
the mean distance $d$ between all oscillator pairs. The distance
between two oscillators is defined as the number of network links
along the path of minimal length joining them. Figure \ref{figure5}
shows the time evolution of $d$ for three different values of the
coupling strength, $r=0.4$, $1.0$ and $2.0$. The three curves are
averages over $100$ different initial conditions. While, in the
considered time span, $d$ grows monotonically for $r=0.4$, the
evolution is manifestly non-monotonic for larger coupling strengths.
At short times, $d$ exhibits a fast growth,  reaches a maximum,  and
then decays slowly. As $r$ increases, the maximum is reached at
shorter times, and the peak becomes sharper. Together with the
observation that asymptotic clustering $C_\infty$ is large, the
small values of $d$ for long times indicate that the network has
evolved, from its initial random structure, to a small-world
pattern.

\begin{figure}
\resizebox{\columnwidth}{!}{\includegraphics*{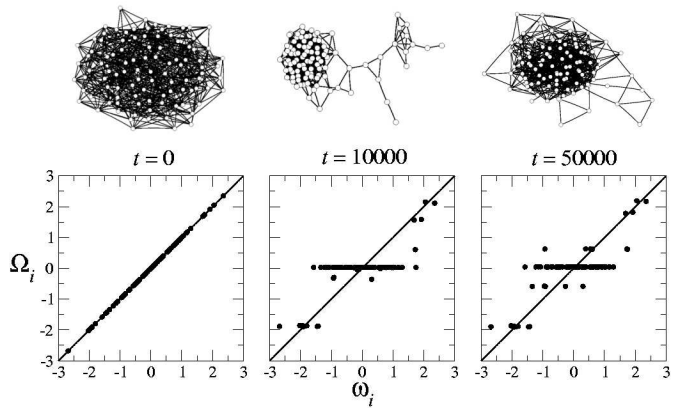}}
\caption{Snapshots of the network structure and the corresponding
average frequency $\Omega_i$  versus natural frequency $\omega_i$
for $r=2.0$, at  three different times, corresponding to the initial
condition, the maximum in the mean distance $d$ (see Fig.
\ref{figure5}), and the stationary state.} \label{figure6}
\end{figure}

Why is it that the mean distance reaches a maximum for intermediate
times? In order to suggest an answer to this question, in Fig.
\ref{figure6} we present snapshots of the network structure,
together with  the corresponding plots of the average frequency
$\Omega_i$ versus natural frequency $\omega_i$, for  $r=2.0$ at
three different times. Networks were drawn using the Pajek software,
optimized for display with the Kamada-Kawai algorithm \cite{Pajek}.
Initially, the network is  random and $\Omega_i = \omega_i$ for all
$i$. For $t=10^4$, the network is clearly divided into clusters,
which naturally leads to a larger value of the mean distance $d$. In
fact, at this time, $d$ reaches its maximum (see Fig.
\ref{figure5}). The plot of average frequencies shows the presence
of well-defined synchronized groups. Inspection of the state of
individual oscillators reveals that there is a direct correspondence
between the groups and network clusters. The largest cluster is
formed by mutually synchronized oscillators with $\Omega_i = 0$,
while oscillators with frequencies far from zero form smaller
clusters. When the system has reached its stationary state, at $t=5
\times 10^4$, six groups of mutually synchronized oscillators can be
clearly distinguished in the plot of average frequencies. The
network structure shows that clusters are now better interconnected,
showing a more compact overall structure. This implies that the mean
distance $d$ must have decreased to a value close to that
corresponding to the initial random network.

Thus, the maximum at intermediate times in the mean distance must be
ascribed to a transient effect in the structural organization of the
network, as groups of synchronized oscillators develop and segregate
from each other. At those times, the network is divided into poorly
interconnected clusters and the mean distance is relatively large.
Later on, further collapse of synchronized groups increases their
interconnection, and $d$ drops. Note that, although counterintuitive
at first sight, this behaviour is not incompatible with the
monotonous time growth of the clustering coefficient.

\begin{figure}
\resizebox{\columnwidth}{!}{\includegraphics*{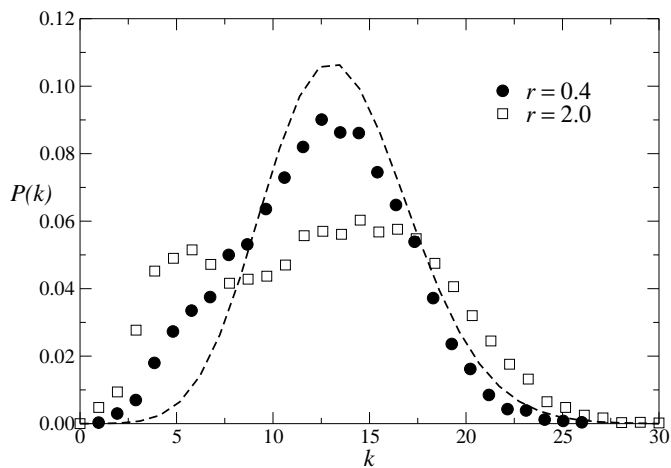}}
\caption{Degree distribution $P(k)$ for two values of the coupling
strength $r$. The dashed curve corresponds to the Poisson
distribution of the random initial network.} \label{figure7}
\end{figure}

Additional characterization of the resulting network structure is
provided by the degree distribution $P(k)$, which gives the
frequency of nodes with exactly $k$ neighbours. We have analyzed the
stationary distribution $P(k)$ for various coupling strengths and,
in Fig. \ref{figure7}, we show results for two representative values
of $r$. For comparison, the Poisson distribution of the random
initial network $p(k)=\exp (-\lambda) \lambda^{k}/k!$, with
$\lambda=12$, is shown as a dashed curve. As $r$ grows, the
distribution deviates from the Poissonian shape. Its maximum
flattens and shifts to the right, while the small-$k$ range becomes
more populated. Eventually, the distribution acquires a bimodal
shape, with a secondary peak at small $k$, as shown in Fig.
\ref{figure7} for $r=2$. Individual analysis of connectivities
reveals that this peak represent the contribution of oscillators in
small synchronized groups, whose frequencies are relatively far from
zero. Large connectivities, on the other hand, belong to oscillators
in the largest group.

\section{Conclusions}

We have presented a model of adaptive evolution in a system formed
by coupled non-identical oscillators. The attractive interaction
between oscillators is described by means of a dynamic network, that
evolves by favouring the interactions between elements whose average
frequencies are similar. We have analyzed in detail the
synchronization properties of the model, the underlying network
structure, and the relation between synchronization and structure.

We have begun our study  by considering the synchronization order
parameter $z$. The analysis of the time evolution of $z$ shows that
it reaches a stationary value that grows as the coupling strength
between oscillators increases. This reveals the existence of
partially synchronized states. In order to obtain more information
on these states, we have inspected snapshots of the mean frequency
versus the natural frequency, for different values of the coupling
strength. This revealed the presence of groups of oscillators which
had become synchronized even for very small values of the coupling
strength.

The presence of several synchronized groups signals to a non-trivial
structure in the underlying network. In order to analyze the network
structure we consider first the clustering coefficient $C$. The time
evolution of $C$ shows that it reaches a stationary value
$C_{\infty}$ that strongly depends on the coupling strength $r$. In
fact,  $C_{\infty }$  presents a maximum at $r_{max} \sim 1$. This
indicates that, as a result of the rewiring process, the network
reaches a more complex structure when the interactions between the
oscillators are neither too weak nor too strong.

It is interesting to compare the behavior of the stationary value of
the clustering with the results obtained with a very similar model
for chaotic maps \cite{Gong}. In that case the stationary value of
$C$ was either zero for small  coupling strength, or saturated close
to $0.6$ for larger values of the coupling strength.

As a next step in the analysis of the network structure, we have
considered the mean distance $d$. The observation that $d$ is small
for long times, together with the fact that the stationary
clustering $C_{\infty}$ is large, show that the network evolves from
a random initial structure to a small-world network. Through an
inspection of individual oscillators we have established a direct
connection between synchronized groups and clusters in the network
structure. This makes it possible to explain the non-monotonic
transient behavior of the mean distance, and the bimodal shape of
the degree distribution for large values of the coupling strength.

It is worth recalling the synchronization phenomenon of flashing
fireflies, as described by J. and E. Buck. They observed that
fireflies brought to their hotel room in Bang\-kok ``...first flew
about...then settled down in small groups''  and finally ``...the
flashing within each group became mutually synchronous''
\cite{Buck2}. The similarity with the behavior observed in our model
suggests that the mechanism of enhancing interactions between
dynamical elements whose internal states are similar plays a key
role in the adaptive emergence of coherent dynamics.

\section{Acknowledgments}

P.M.G. acknowledges financial support from Consejo Nacional de
Investigaciones Cient\'{\i}ficas y T\'ecnicas CONICET
(Argentina), Fundaci\'on Antorchas (Argentina), ANPCYT PICT2003
(Argentina) and  ICTP grant NET-61 (Italy).

\end{document}